\documentclass[aps,prl,superscriptaddress,twocolumn]{revtex4-1}
\usepackage{graphicx}
\usepackage{amssymb}
\usepackage{amsmath}
\usepackage{multirow}
\usepackage{float}
\graphicspath{{images/}}
\usepackage{float}

\usepackage{changes}

\definechangesauthor[name={Gonzalo Medina}, color=orange]{GM}
\definechangesauthor[name={John Doe}, color=blue]{JD}

\colorlet{Changes@Color}{blue}

\begin{document}

\title{Spontaneous superconducting diode effect in non-magnetic Nb/Ru/Sr$_2$RuO$_4$ topological junctions}

\author{Muhammad~Shahbaz~Anwar}\email[E-mail:]{m.s.anwar476@gmail.com} \affiliation{Department of Physics, Kyoto University, Kyoto 606-8502, Japan}\affiliation{Department of Materials Science \& Metallurgy, University of  Cambridge, CB3 0FS Cambridge, United Kingdom}

\author{Taketomo~Nakamura} \affiliation{Department of Physics, Kyoto University, Kyoto 606-8502, Japan} \affiliation{Fukuoka Institute of Technology, 3-30-1 Wajiro-higashi, Higashi-ku, Fukuoka, 811-0295 Japan}

\author{Ryosuk~Ishiguro} \affiliation{Department of Mathematical and Physical Sciences, Faculty of Science, Japan Women's University, Tokyo 112-8681, Japan} \affiliation{Department of Applied Physics, Faculty of Science, Tokyo University of Science, Katsushika  Tokyo 162-8601, Japan}

\author{Shafaq~Arif} \affiliation{Department of Physics, Lahore College for Women University, Lahore, 54000, Pakistan}

\author{Jason~W.~A.~Robinson} \affiliation{Department of Materials Science \& Metallurgy, University of  Cambridge, CB3 0FS Cambridge, United Kingdom}

\author{Shingo~Yonezawa} \affiliation{Department of Physics, Kyoto University, Kyoto 606-8502, Japan}

\author{Manfred~Sigrist} \affiliation{Theoretische Physik, ETH Zurich, CH-8093 Zurich, Switzerland}

\author{Yoshiteru~Maeno} \affiliation{Department of Physics, Kyoto University, Kyoto 606-8502, Japan}\affiliation{Toyota Riken–-Kyoto University Research Center (TRiKUC), Kyoto 606-8501, Japan}

\maketitle

\noindent { \Large \bf Abstract}

\noindent
{\bf Non-reciprocal electronic transport in a material occurs if both time reversal and inversion symmetries are broken. The superconducting diode effect (SDE) is an exotic manifestation of this type of behavior where the critical current for positive and negative currents are mismatched, as recently observed in some non-centrosymmetric superconductors with a magnetic field. Here, we demonstrate a SDE in non-magnetic Nb/Ru/Sr$_2$RuO$_4$ Josephson junctions without applying an external magnetic field. The cooling history dependence of the SDE suggests that time-reversal symmetry is intrinsically broken by the superconducting phase of Sr$_2$RuO$_4$. Applied magnetic fields modify the SDE dynamically by randomly changing the sign of the non-reciprocity. We propose a model for such a topological junction with a conventional superconductor surrounded by a chiral superconductor with broken time reversal symmetry.}

\vspace{5mm}

\noindent { \Large \bf Introduction}

\noindent Diodes are one-way electronic switches for charge flow, and are a fundamental component in modern circuits. Recently a superconducting diode effect (SDE) (i.e. non-reciprocal flow of supercurrents) was observed in superconducting superlattices~\cite{Ando2020,Narita2022}, 2-dimensional materials~\cite{Ye2022,Bauriedl2022,Wakatsuki2017}, multilayers~\cite{Hou2022} and Josephson junctions~\cite{Baumgartner2022,Wu2022,Pal2021}. Such a dissipationless SDE paves-the-way to the development of energy-efficient electronics and computation. 

Broken symmetries in materials can produce new quantum phenomena including magnetochiral anisotropy (MCA). On a microscopic level the lack of inversion symmetry induces specific forms of spin-orbit coupling (SOC). For instance, MCA in a chiral conductor is described through the non-linear dependence of the resistivity $R(H)=R_{\rm o}(1-\gamma H I)$ for specific current directions~\cite{Rikken1997,Rikken2001}, where $\gamma$ is the MCA coefficient originating from SOC, $H$ is magnetic field applied perpendicular to the current flow and $R_{\rm o}$ is linear resistance. 

Non-reciprocity in superconductors is seen as a difference in critical current for positive and negative flow of the current, constituting a SDE. A recently discussed example are non-centrosymmetric superconductors, where a lack of an inversion center in the crystal lattice induces antisymmetric SOC and influences the superconducting phase. Applying a magnetic field to a non-centrosymmetric superconductor can give rise to SDE. This effect is seen for systems that exhibit Rashba-type SOC due to the absence of specific mirror symmetries in crystal lattice. A magnetic field directed along such planes establishes a helical phase with non-reciprocal in-plane supercurrents perpendicular to the magnetic field~\cite{Daido2022,Yuan2022,Ilic2022,Scammell2022,He2022}. An example includes Nb/Ta/V superlattices with antisymmetric SOC; the SDE in such structures is observed when the magnetic field is applied perpendicular to the current flow, such that both inversion and time reversal symmetries are broken. Insertion of ferromagnetic layers is reported to serve a similar role as external magnetic field~\cite{Narita2022}. In case of a magnetic field-free SDE~\cite{Narita2022,Wu2022,Jeon2022,Lin2022,Merida2023}, broken inversion symmetry at the interfaces may produce required SOC and presence of a magnetic layer in the devices breaks time-reversal symmetry. This may also be true for EuS/Nb/Pt multilayers~\cite{Hou2022}.

The SDE can also occur without an applied magnetic field in a superconductor with spontaneously broken time-reversal symmetry and, as noted in~\cite{Zinkl2022}, without SOC if inversion symmetry is violated. One of the first observations of this type of ``spontaneous'' SDE was reported for Sr$_2$RuO$_4$ (SRO) but above the bulk superconducting transition temperature where a filamentary superconducting phase so-called 3-Kelvin (3-K) phase exists~\cite{Hooper2004}.

This article investigates Josephson junctions based on SRO, which exhibit an unconventional superconducting order parameter with spontaneous time-reversal symmetry violation~\cite{Maeno2012,Mackenzie2017,Anwar2021}. A eutectic single crystal of SRO with Ru metal inclusions shows two phases of superconductivity, bulk with a transition temperature ($T_{\rm c}$) of 1.5~K (1.5-K phase) and a filamentary superconducting phase at the Ru/SRO interface~\cite{Ghosh2017} below 3~K (the so-called 3-K phase~\cite{Maeno1998,Haka2009}). It has been speculated that strain at the interface may be responsible for the locally enhanced onset of superconductivity in the 3-K phase~\cite{Hicks2014,Taniguchi2015,Hicks2017}. Bulk superconductivity in SRO is sensitive to impurities, consistent with an unconventional order parameter~\cite{Mackenzie1998}. Evidence for broken time-reversal symmetry in the superconducting state originates from the observation of an intrinsic magnetic field below $T_{\rm c}$ by zero-field muon spin relaxation~\cite{Luke1998,Grinenko2021} and Kerr rotation~\cite{Xia2006} experiments. Furthermore, electronic transport measurements of SRO junctions indicate two-fold degenerate domains~\cite{Kindwingira2006,Nakamura2011,Nakamura2012,Anwar2013,Anwar2017}. Among other experiments, these observations suggest a two-component order parameter with chiral symmetry, while the symmetry of the electron pairing is still under debate~\cite{Hassinger2017, Romer2019,Pustogow2019,Ishida2020,Petsch2020,Kivelson2020,Suh2020,Li2021,Grinenko2021}.      

In this article, we report a magnetic field-free (spontaneous) SDE in Nb/Ru/SRO Josephson junctions. Josephson currents are observed below about 2~K with the SDE appearing below 1.4~K when bulk superconductivity is established in SRO. The difference between the critical current for opposite current directions increases below this onset and passes through a maximum around 0.5~K, below which the Ru-metal inclusion becomes intrinsically superconducting~\cite{Nago2014}. This SDE depends on thermal cycles and is affected by applying magnetic fields parallel to the basal plane of SRO. 

 \begin{figure*}
\includegraphics[width=17cm]{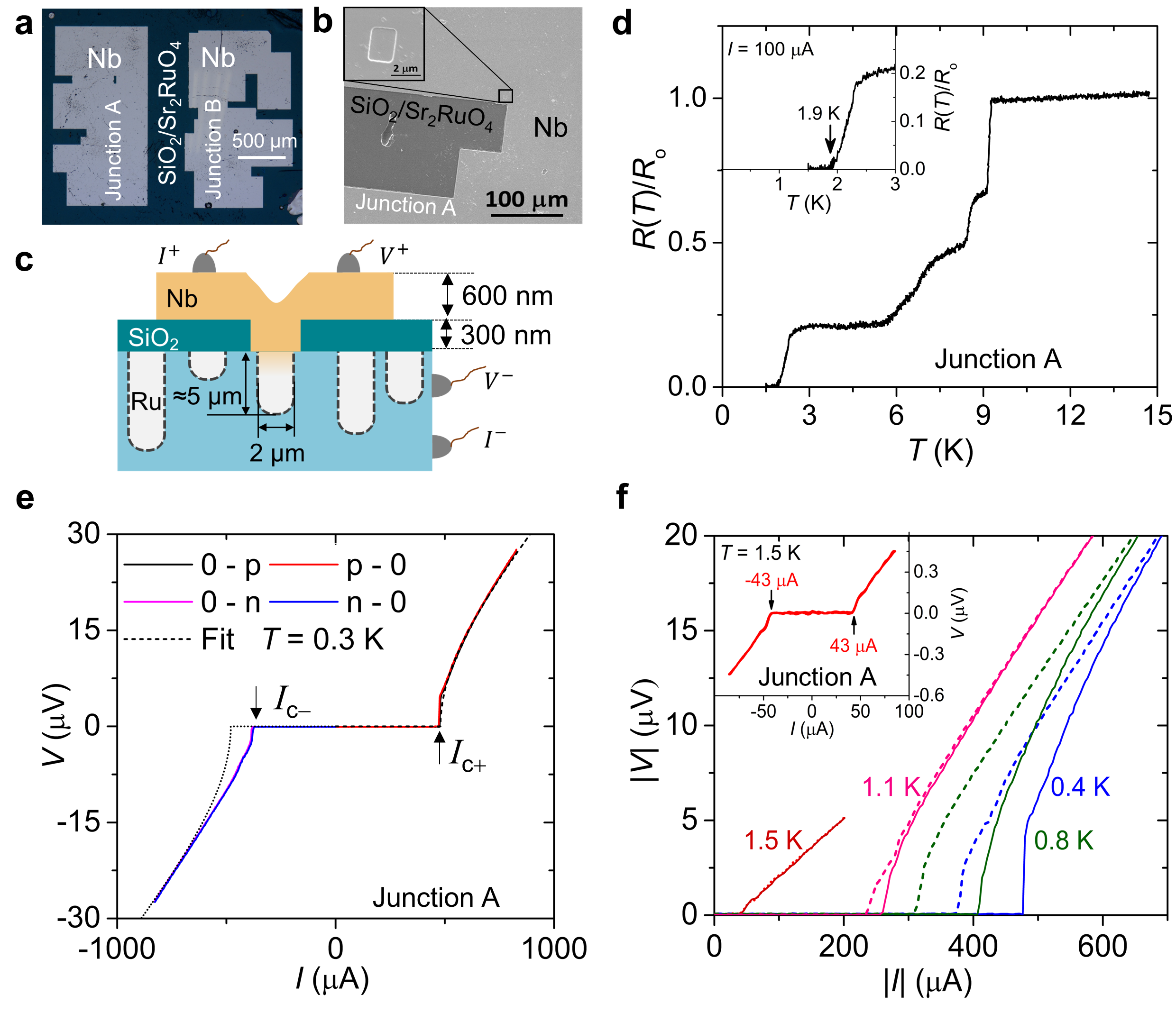}
\caption{{\bf Junction geometry and electronic transport properties}. {\bf a} A scanning electron micrograph showing two junctions A (Left) and B (right). The brighter areas are the two top Nb electrodes and the darker area is the SiO$_2$/SRO substrate. {\bf b} The SiO$_2$ insulating layer is removed in small area covering the top of a single Ru inclusion as shown in the inset (junction A). An analogous setup is realized for the junction B. {\bf c} A schematic view of a Nb/Ru/SRO junction fabricated by depositing 600 nm thick Nb layer on top of a SiO$_2$ layer with a hole allowing direct contact of Nb to a single Ru inclusion of SRO-Ru eutectic crystal (layer orientation with normal vector along the $c$-axis of SRO). {\bf d} Resistance as a function of temperature of junction A. The first transition at 9.2~K and the transition at 2.2~K correspond to the onset of superconductivity in the Nb electrode and in the SRO-Ru eutectic crystal, respectively. Zero resistance is achieved at 1.7~K well above 1.5~K that indicates the Josephson coupling develops first through 3-K phase. There are two more transition at 8.5~K and 6.5~K, which corresponds to the geometry of the junction. Inset shows the $R$($T$) at lower temperature. {\bf e} A non-hysteretic $V$($I$) curve measured at 0.3~K with current scans in the sequence of i) zero to positive (0~-~p), ii) positive to zero (p~-~0), iii) zero to negative (0~-~n) and iv) negative to zero (n~-~0). The comparison with a symmetric theoretical $V$($I$) curve (black dotted line) clearly shows non-reciprocity of the experimental junction: $I_{\rm c+} \neq |I_{\rm c-}|$. {\bf f} $V$($I$) curves for the absolute values of current and voltage at different temperatures. Solid and dotted lines mark the branches belonging to positive and negative current direction, respectively. The difference between magnitudes of $I_{\rm c+} $ and $ I_{\rm c-}$ decreases with increasing temperature and disappears above 1.4~K (see also inset with an $V$($I$) curve at 1.5~K).}
\label{device-RT}
\end{figure*}

\begin{figure*}
\includegraphics[width=17.2cm]{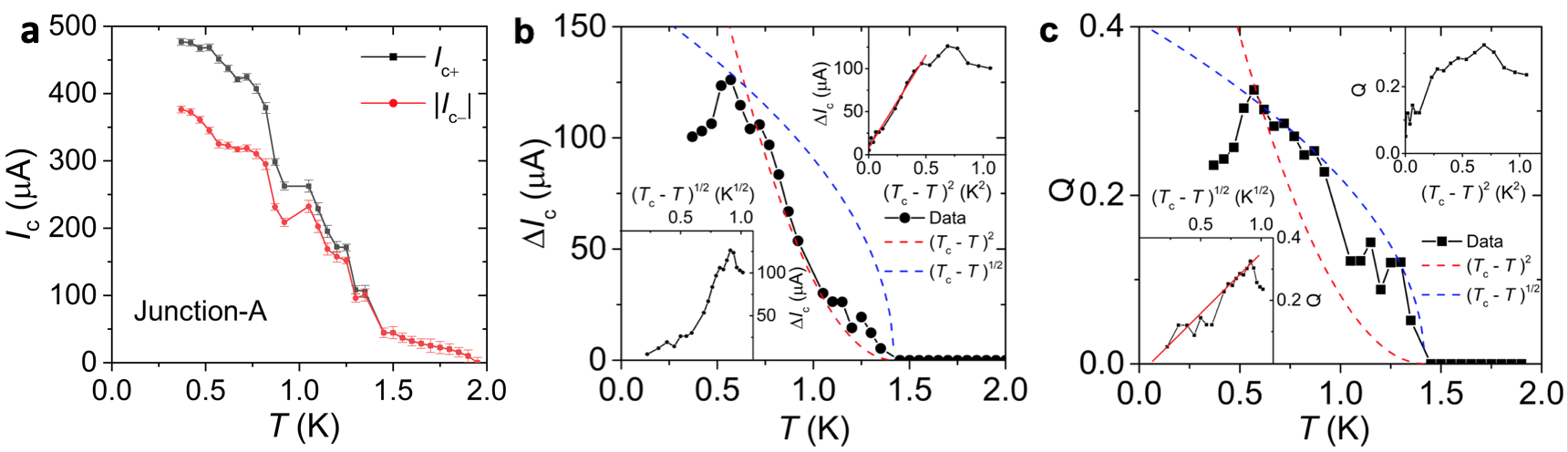}
\caption{{\bf Temperature dependence of the superconducting diode effect}. {\bf a} Critical currents $I_{\rm c+}$ and $|I_{\rm c-}|$ vs $T$. The two critical currents are non-vanishing below 2~K as a result of the 3~K-phase and start to differ from each other below $ T_{\rm c} \approx $ 1.4~K when SRO shows bulk superconductivity.  Above $T_{\rm c}$ there is no difference in the critical currents. $\Delta I_{\rm c} = I_{\rm c+} -|I_{\rm c-}|$ and $Q$ defined as $ (I_{\rm c+} - |I_{\rm c-}|)/(I_{\rm c+} + |I_{\rm c-}|)$ are plotted against $T$ in panel {\bf b} and {\bf c}, respectively. The insets in both panels show a distinct $T$-dependence close to the onset of the bulk $T_{\rm c}$, which is $(T_{\rm c} - T)^2 $ for $ \Delta I_{\rm c}(T) $ and $(T_{\rm c} - T)^{1/2} $ for $Q(T) $. Note that the standard deviation in $I_{\rm c}$ for all temperatures is smaller than the data points.}
\label{temp}
\end{figure*}

\begin{figure*}
\includegraphics[width=17.2cm]{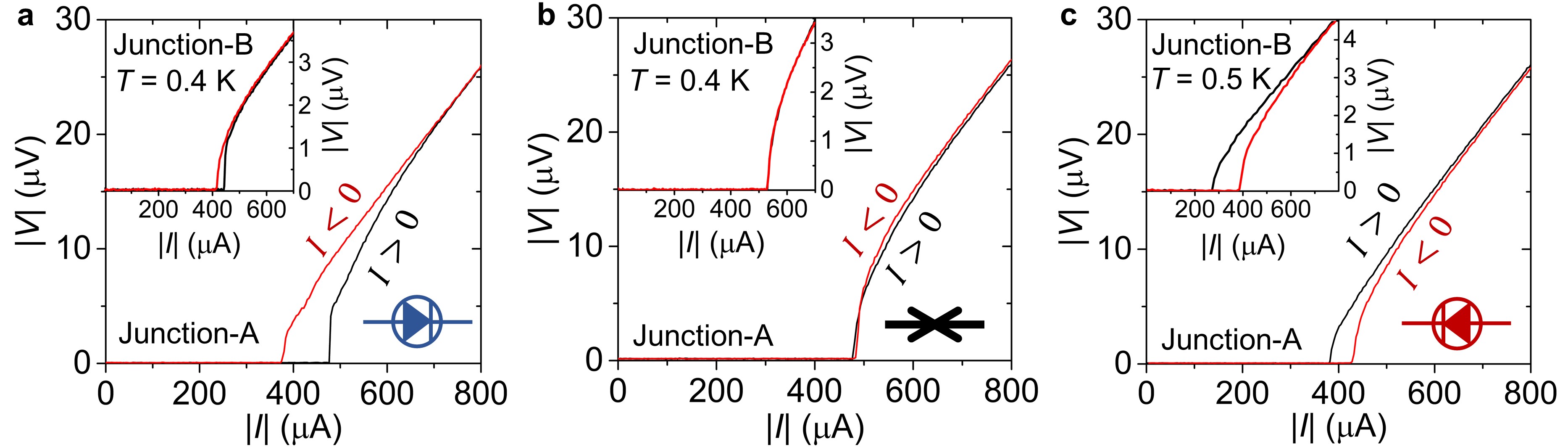}
\caption{{\bf Effect of thermal cycling on the superconducting diode effect}. Three different $V$($I$) curves with absolute values of current and voltage measured for different cooling cycles; the black (red) curve shows the positive (negative) branch. {\bf a} Junction A exhibits a forward diode effect with $I_{\rm c+} > I_{\rm c-}$, {\bf b} standard junction behavior with $I_{\rm c+} = |I_{\rm c-}|$ and {\bf c} reverse diode effect with $I_{\rm c+} < |I_{\rm c-}|$. Insets show $V$($I$) curves for junction B for equivalent thermal cycles.}
\label{coolingcycle}
\end{figure*}

\begin{figure*}
\includegraphics[width=17cm]{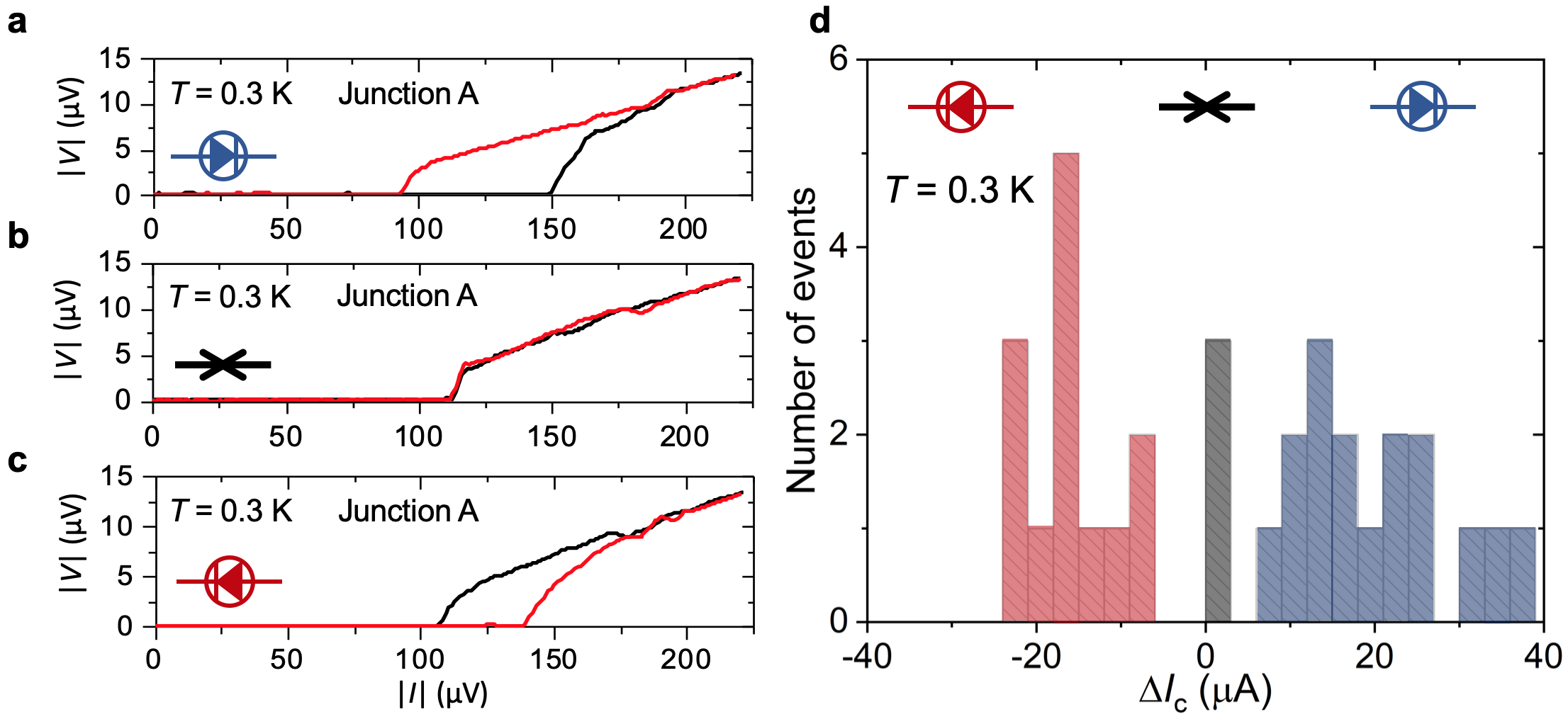}
\caption{{\bf Effect of in-plane magnetic fields on the superconducting diode effect}. The effect of magnetic fields applied perpendicular to the $c$-axis is investigated for junction-A using the following sequence: i) the magnetic field is applied in the superconducting state (zero-field cooled state), ii) the magnetic field is switched off at 0.3~K after measuring some $V$($I$) curves from 0 to $+$20 mT, back to $-$20 mT and back to zero and iii) $V$($I$) curves are measured in the zero field at the same temperature.  The $V$($I$) characteristic behave unstable as long as the magnetic field is switched on and does not allow to determine critical currents reliably (see Figure~S1c~and~S1d). After switching off the magnetic field $V$($I$) curves show relatively stable critical currents, but the non-reciprocal component $\Delta I_{\rm c}$ is ``random''; $V$($I$) scans at different times sometimes show the behavior of {\bf a} forward diode $I_{\rm c+}>|I_{\rm c-}|$, {\bf b} standard junction $(I_{\rm c+} \approx |I_{\rm c-}|)$ and {\bf c} reverse diode $(I_{\rm c+}<|I_{\rm c-}|)$. {\bf d} Histograms of $\Delta I_{\rm c}$ for 34 sequential $V$($I$) scans showing quantitative scatter of the SDE around $\Delta I_{\rm c}=\pm$20~$\mu$A. Blue, black and red bars indicate forward diode effect ($\Delta I_{\rm c} > 0$), standard junction ($\Delta I_{\rm c} = 0$) and reverse diode effect ($\Delta I_{\rm c} < 0$). Note that $V$($I$) curves are measured successively at 0.3~K with a negligible time delay between measurements.}
\label{field}
\end{figure*}

\vspace{5mm}
\noindent { \Large \bf Results}

\vspace{2mm}
\noindent {\bf Junction fabrication and electronic characteristics}
\noindent Nb/Ru/SRO junctions were fabricated by depositing a 600-nm-thick Nb film on the $ab$-surface of the SRO-Ru eutectic crystal covered by an insulating mask of SiO$_2$ with a
hole such that the Nb film is in contact with a single Ru-inclusion only (see Fig.~\ref{device-RT}~a-c); details are given in the Methods Section. The $\mu$m-sized Ru-metal inclusion is embedded in a single crystal of SRO such that the interface between the two forms a Josephson junction. Assuming a conventional $s$-wave superconducting phase within Ru (proximity-induced by Nb and intrinsic below 0.5~K) and a chiral superconducting phase in SRO, this Josephson junction behaves as a topological junction as it involves a winding of the Josephson phase~\cite{Nakamura2011,Nakamura2012,Anwar2013,Anwar2017,Kaneyasu2010-1,Kaneyasu2010-2,Etter2014}. We have measured electronic transport properties through Nb/Ru/SRO junctions down to 0.3~K (electronic setup is described in the Methods Section). Figure~\ref{device-RT}d shows the resistance vs temperature $R$($T$) measured with a 100~$\mu$A dc current. Note that the $R$ was measured in delta mode and the average value of $R$ was measured with step-like current variations (positive and negative) with a frequency of 5~Hz. The drop of resistance coincides with the onset of superconductivity in Nb at $\approx$ 9.2~K, followed by a further transition around 2.2~K corresponding to the 3-K phase at the Ru/SRO interface. Zero-resistance is achieved below 1.7~K which is above the bulk $T_{\rm c}$ of 1.4~K of SRO. Note that we observe some additional transitions between 9.2~K ($T_{\rm c}$ of Nb electrode) and 2.2~K (onset $T_{\rm c}$ of SRO-Ru) that may relate with the geometry of the junctions (narrow parts at the corners of the junction, wider pads and leads). Such transitions are common in these types of junctions.

Note that SRO single crystal is a good metal. At 4~K it exhibits resistivity of 1$\mu \Omega$cm along the $ab$-surface~\cite{Anwar2015}, which yields the resistance of $\approx$~10~$\mu \Omega$. In the normal state of the junctions (above 2.2~K) the resistance of the single crystal is ignorable compared with the junction resistance at 4~K (30 m$\Omega$ for junction A and 7 m$\Omega$ for junction B). This means for electron transport the junction resistance is mainly dominating than that of the single crystal. On the other hand, we utilized Ru-SRO eutectic crystal to fabricate the junctions. The eutectic single crystals contain a number of Ru inclusions and a 3K phase develops around the inclusions. A network of Ru inclusions can provide a connection through 3K phase which can lead to the zero resistance V-lead.

\vspace{2mm}
\noindent {\bf Temperature effect on non-reciprocity}

\noindent To investigate non-reciprocity of the critical current we measured current voltage $V$($I$) characteristics between 0.3 and 2~K. Figure~\ref{device-RT}e shows a set of $V$($I$) curves at 0.3~K using the following current cycles: i) zero to positive (0~-~p), positive to zero (p~-~0), zero to negative (0~-~n) and negative to zero (n~-~0). As there is no hysteretic behavior visible in the $V$($I$) curves, the junction is in the overdamped regime. A fit based on $V = R(I^2 - I_{\rm c}^2)^{1/2}$ (dotted line) shows that the critical currents  $I_{c+}$ (positive direction) is higher than $|I_{\rm c-}|$ (negative direction).  
To compare opposite current directions for different temperatures, in Fig.~\ref{device-RT}f we have plotted $V$($I$) curves for absolute currents and voltages, where solid (dotted) lines indicate the positive (negative) branch of the curve. At $ T = 1.5$~K there is no difference between $I_{\rm c+}$ and $|I_{c-}|$ (see also the inset of Fig.~\ref{device-RT}f), $\Delta I_{c} = I_{\rm c+} - |I_{\rm c-}|$ is non-vanishing at lower temperatures. Similar behavior is observed for junction-B (see Figure~S2).

In Fig.~\ref{temp}a-c we have plotted the $T$-dependence of the critical current for both directions and their difference. Both $I_{c+}$ (black squares) and $|I_{\rm c-}|$ (red circles) decrease for increasing temperature and merge above 1.4~K (Fig.~\ref{temp}a). The SDE displayed in $\Delta I_{\rm c}$ grows continuously below 1.4~K and shows a maximum around 0.5~K (Fig.~\ref{temp}b). The $T$-dependence below $ T_{\rm c} $ follows  $\Delta I_{\rm c} (T) \propto (T_{\rm c} - T )^2 $, similar to the theoretical expectations for non-centrosymmetric superconductors~\cite{Daido2022,He2022}. 

The efficiency of the diode effect can be scaled with the quality parameter $Q$, defined as $Q = \frac{I_{c+} - |I_{c-}|}{I_{c+} + |I_{c-}|}$. For our junctions, we find an increase in the $Q$ with decreasing temperature, following $\sqrt{T_{\rm c} - T}$ (Fig.~\ref{temp}c), which is an expected behavior~\cite{Daido2022,He2022}. The $Q$ reaches a maximum of 0.34 (34$\%$) at 0.5~K.

\vspace{2mm}
\noindent {\bf Effect of thermal cycle on non-reciprocity}

\noindent After cooling to 0.3~K, the junctions are warmed to 10~K to reach the normal state. The junctions are then cooled to $\approx$ 3~K and maintained at this temperature for 30~min before cooling to 0.3~K. We observe that with each cooling cycle our junctions randomly change the SDE state among one of the three states, forward diode ($\Delta I_{\rm c}= +$; see Fig.~\ref{coolingcycle}a), standard junction ($\Delta I_{\rm c}= 0$; see Fig.~\ref{coolingcycle}b) and reverse diode ($\Delta I_{\rm c}= -$; see Fig.~\ref{coolingcycle}c). The junction remains in its state persistently down to the lowest temperature and at a fixed temperature within one thermal cycle $\Delta{I_{\rm c}}$ remains almost constant (see Figure~S3). Note that the standard junction behaviour ($\Delta I_{\rm c}=0$) occurs rarely. For 12 thermal cycles, our junctions show about 60\% forward diode, 30\% reverse diode and 10\% standard junction behaviour. Thus, we conclude that the SDE is dependent on the cooling cycle, a feature not seen in non-centrosymmetric superconductors where the sign of non-reciprocity depends on the direction of the applied magnetic field. 

\vspace{2mm}

\noindent {\bf Effect of magnetic fields on non-reciprocity}

\noindent We note that the SDE in our junctions does not require an external magnetic field nor the addition of a magnetic layer. Nevertheless, we have investigated the effect of magnetic fields oriented parallel to the $ab$-surface of SRO on the SDE. Such magnetic fields cause $I_{\rm c}$ of the junctions to become unstable (see Figure~S6). After switching off the magnetic field the junctions $I_{\rm c}$ are partially re-stabilized but the magnitude and sign of the SDE is randomly changing. Figure~\ref{field}a-c shows the $V$($I$) curves in zero magnetic field measured after an in-plane field cycle ($\pm$20mT) at 0.3~K, showing that all three states (defined above) are reached. Considering a number of field sweeps, Fig.~\ref{field}d displays histograms of $\Delta I_{\rm c}$ at zero magnetic field. This demonstrates a dynamical behavior stimulated by the external magnetic field. We would like to emphasize the $ab$-plane direction of the magnetic field would not couple to the chirality of the superconducting order parameter in SRO, thus, would unlikely reverse chirality. Similar behavior has been observed in junction B (see Figure~S4).

\begin{figure*}
\includegraphics[width=17cm]{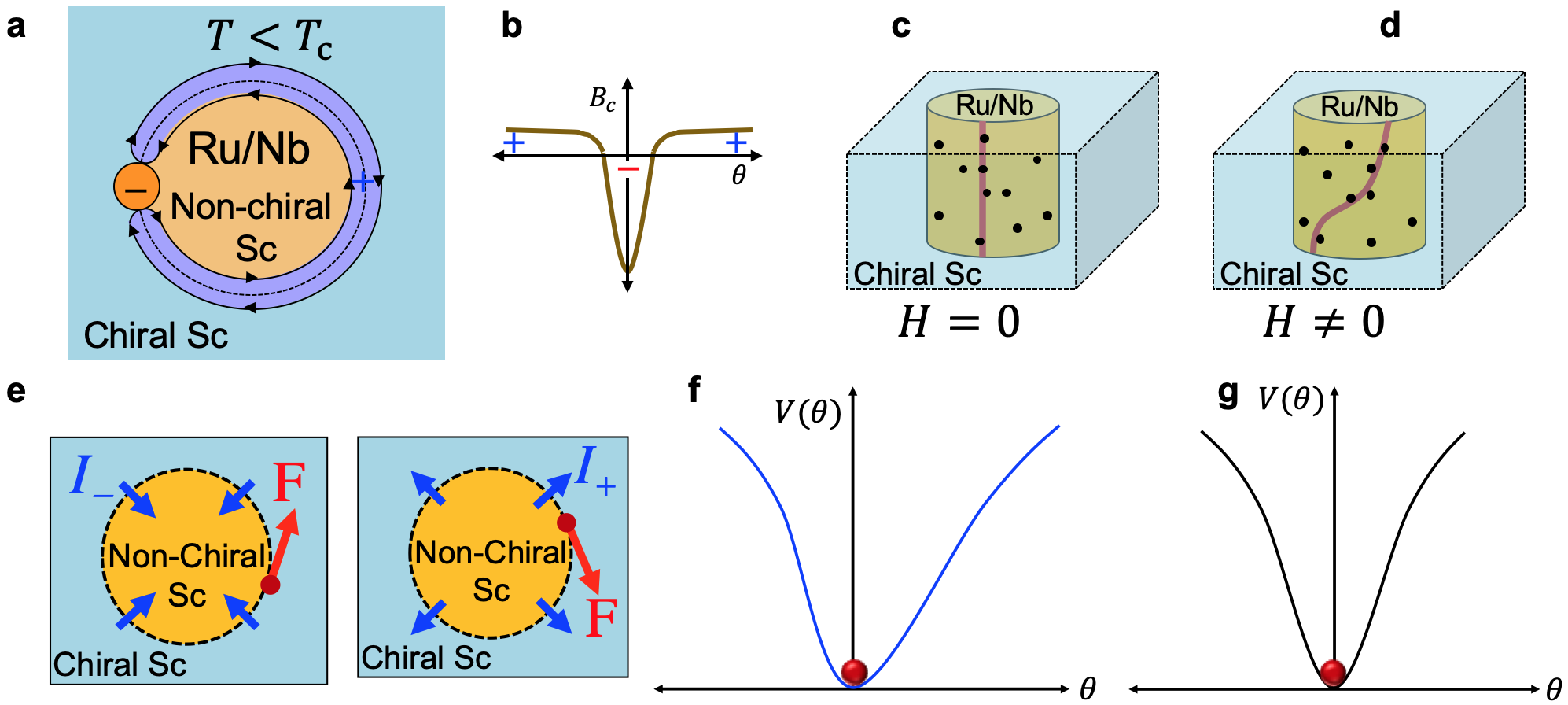}
\caption{{\bf Spontaneously induced vortex at the Ru/SRO interface}. {\bf a} A schematic illustration of the concentrated flux line (vortex) at the interface between a non-chiral superconductor (Sc) and chiral Sc, which is Ru/SRO interface for our junctions. This appears below bulk superconducting transition temperature of SRO. The magnetic flux of this vortex is compensated by a magnetic flux in opposite direction spread out over the remaining interface (here for a cylindrically shaped Ru-inclusion embedded in SRO). {\bf b} Flux distribution, where $\theta$ is the angle indicating the position on the interface around the inclusion. {\bf c} Illustration of pinning of the flux line at scattered pinning sites in zero field and {\bf d} the flux line tilted under applied in-plane magnetic field. {\bf e} Schematic picture of Lorentz force on the flux line (red dot) for negative and positive current direction. Note that the force direction reverses for the superconducting phase of opposite chirality in SRO. {\bf f} Asymmetric pinning potential for the vortex which allows for easier depinning for motion to the right, leading to forward bias diode effect. {\bf g} Symmetric pinning potential that has no preferred depinning direction. There is no diode effect.}
\label{theory}
\end{figure*}

\vspace{5mm}

\noindent {\Large \bf Discussion}

\noindent
First, we discuss whether our results are intrinsic to the junctions or related to the junction geometry. One might suspect that the observed SDE is spurious due to structural artefacts in the junction such as crystallographic defects inhomogeneous interfaces and magnetic impurities that may lead to anisotropic $V$($I$) curves mimicking a SDE. However, effects triggered in this way are unlikely to display a history dependence with cooling cycles, but would rather remain persistent. This is clearly not the case for the SDE observed in our junctions. The temperature effect and dynamical behavior of SDE under magnetic fields suggest that the SDE is a feature connected with the non-trivial superconducting state of SRO. 

Our junctions consist of non-magnetic materials (see Figure S7), apart from possible local magnetism of SRO for $c$-axis surface attributed to an in-plane rotation of the RuO$_6$ \cite{Matzdorf2000,Angelo2021}, which may also yield broken time reversal symmetry locally. According to recent experimental studies the surface magnetism appears already at temperatures below 50~K~\cite{Angelo2021}, much higher than the onset $T_{\rm c}$ of the 3K phase. Thus, it is very unlikely that the sharp onset of the SDE at 1.4~K (SRO bulk $T_{\rm c} =$ 1.4~K for the single crystal used to fabricate the devices) is induced by the kind of surface magnetism. Thus, the weak surface magnetism may not play any significate role in emerging such a strong SDE in our junctions.

It has been demonstrated that vortices induced by applied and/or residual magnetic fields may lead to the non-reciprocal electronic transport in a superconducting device, which can mimic the SDE \cite{Hou2022,Chahid2023,Gutfreund2023}. We believe that this is not the case for our devices, as we observe SDE in a virgin state as well; a state before applying any magnetic field and the devices are measured in a cryostat shielded with $\mu$-metal. On the other hand, vortices induced because of a residual field of a superconductor magnet coil should be persistent particularly at low temperatures. Therefore, SDE originated from such a residual field must not change the sign with a cooling cycle happening below 10~K. In this temperature range only flux density may change but not the sign/direction \cite{Romanenko2014,Kubo2016}.
 
We believe that the understanding of the non-reciprocal behavior of these junctions is connected with the mechanism limiting the Josephson critical current between the Ru-inclusion embedded in bulk SRO and the bulk SRO. The Ru-SRO interface has been found to be atomically sharp and providing good conditions for Josephson coupling between two superconductors with order parameters of different symmetry~\cite{Ghosh2017}. 

Through the Nb top electrode the Ru inclusion acquires conventional $s$-wave superconductivity due to the proximity effect and becomes intrinsically superconducting below $T=$ 0.5~K. We assume that the bulk superconductivity (1.5-K phase) of SRO has a chiral order parameter whose phase varies under rotation around the crystal $c$-axis. Under this condition, the in-plane Josephson coupling between Ru and SRO is phase frustrated on a closed interface like for an inclusion of cylindrical geometry (closed surface), because the $s$-wave order parameter of Ru would interact with a different phase for the chiral order parameter for different junction orientation~\cite{Kaneyasu2010-1,Etter2014}. Around the closed interface the Josephson phase picks up a full $2\pi$-phase winding going once around the interface between Ru-SRO enclosing the $c$-axis. It has been shown that this configuration induces a spontaneous magnetic flux distribution ($c$-axis oriented) along the interface with one rather well localized vortex-like flux line in order to relax that phase frustration~\cite{Kaneyasu2010-1}, as illustrated in Fig.~\ref{theory}a-c. Note that the total magnetic flux of the interface vanishes, compensated by widely spread flux of opposite sign (Fig.~\ref{theory}a and b).  When a current flows through the interface a Lorentz force acts on the vortex pushing it along the interface (Fig.~\ref{theory}e and f). Thus, such a current only flows without dissipation, as long as this vortex remains pinned and does not move. It was shown theoretically that the depinning of this spontaneous vortex would determine the limiting of the Josephson current and defines the critical current~\cite{Etter2014}. 
In case of a perfectly symmetric pinning potential (Fig.~\ref{theory}g) to move the vortex left or right the depinning force has the same magnitude for both directions of the current, which leads to the standard junction behavior. In reality pinning potentials around a Ru inclusion are generally asymmetric simply by the nature of pinning defects and the irregular geometry of the inclusion. Consequently, we find different limiting (critical) current for flow direction into and out of the Ru-inclusion (Fig.~\ref{theory}e and f). Thus, the origin of the non-reciprocal behavior in this case relies on broken time-reversal symmetry (chiral order parameter in SRO) responsible for existence of the spontaneous flux line and the broken inversion symmetry for the asymmetric pinning potential. Unlike in non-centrosymmetric superconductors the lack of inversion symmetry is not a microscopic feature and, thus, not related to SOC, but is connected with the lack of certain reflection symmetries of the junction~\cite{Zinkl2022}.

It has been demonstrated that the $T_{\rm c}$ of bulk SRO is only enhanced for the uniaxial strain along the [100] direction~\cite{Hicks2014}. This indicates that $T_{\rm c}$ of the emerged 3-K phase around a Ru in SRO-Ru eutectic can be distributed non-uniformly around the Ru inclusion. This non-uniform local $T_{\rm c}$ around the Ru-inclusion do not effect our model as SDE is observed only below bulk superconducting transition temperature (in the 1.5-K phase). In our model, we mainly consider the interaction of bulk superconductivity of SRO and induced $s$-wave superconductivity in a Ru inclusion. Therefore, this model is valid for our experimental results. 

The mechanism proposed here yields a behavior which can depend on thermal cycling, because chirality is a feature of spontaneous symmetry breaking and can have either sign equally likely. For opposite signs the flux line would have opposite magnetic flux reversing the Lorentz force and the sign of the difference of the critical currents, i.e. $ \Delta I_{\rm c} \lessgtr 0 $. Because there may be several pinning positions along the interface, there can also be a history dependence of the magnitude of $ \Delta I_{\rm c}$ besides its sign. 

In this context it is also not surprising that pinning conditions change, when the superconductivity in Ru becomes intrinsic below 0.5~K. It is, however, not obvious why this would always lead to a reduction of non-reciprocity as observed in the experiment. In-plane magnetic fields would tilt or twist the spontaneous flux lines, which likely leads to a change of the pinning conditions as well for the flux line (Fig.~\ref{theory}d). The pronounced influence of the in-plane magnetic field is not astonishing. A tilted flux line may remain trapped in a metastable situation once the magnetic field is turned off. Thus, the relaxation behavior of the flux line returning towards a more $c$-axis oriented position would naturally impact the depinning properties for a certain time after removing the magnetic field. 

As a control experiment to test our model of vortex pinning potential, we fabricated additional junctions in different configurations. A parallel junction, where two parallel Ru inclusions are used to create Nb/Ru/SRO/Ru/Nb Josephson junction (see Figure~S5a). This junction shows an SDE. However, the SDE is not observed in a tip junction where Nb is deposited only on half of the Ru inclusion as shown in Figure~S5b. For a tip junction, 2$\pi$-phase winding cannot be established because of half of the Ru inclusion. A topological junction is not formed which is required to induce a spontaneous vortex state at Ru/SRO interface. Therefore, this shows standard behaviour.

The sign of the non-reciprocity may not only be determined solely by the direction of the chirality. Our model explains that the non-reciprocity may have either $+$ or $-$ sign for a fixed chirality as it depends also on the shape of the pinning potential (see Fig.~\ref{theory}). Further experimental and theoretical studies are required to establish the decisive relation of SDE and chirality including also the formation of chiral domains and the chiral domain wall motion.

In summary, we have demonstrated the existence of a field-free superconducting diode effect in non-magnetic Nb/Ru/SRO topological Josephson junctions prepared by using a SRO-Ru eutectic single crystal. Non-reciprocal component $\Delta I_{\rm c}$, the difference of the critical current in opposite directions, is observed below 1.4~K, when the bulk superconducting phase of SRO is established. $\Delta I_{\rm c}$ rises with lowering temperature and passes through a maximum at around 0.5~K. The sign of $\Delta I_{\rm c}$ varies with different cooling cycles and applied in-plane magnetic field induces an increased variability in $\Delta I_{\rm c}$. These effects are attributed to the chiral superconductivity in the bulk SRO. The observed non-reciprocal behavior is explained by a limiting mechanism of the supercurrent in the topological Josephson junction with a phase frustration. Such a junction is established between the two superconductors with incompatible order parameters: a conventional (non-chiral) in Nb/Ru and a chiral in bulk SRO. The fact that the chiral superconducting phase spontaneously breaks time reversal symmetry and the absence of inversion symmetry due to the geometry of the interface are sufficient conditions to enable the SDE in this type of junctions. This means that no magnetic fields need to be applied and the SDE is a signature of spontaneously broken time reversal symmetry in at least one of the two superconductors. For this reason, the observation of a spontaneous SDE may be a diagnostic tool to detect time reversal symmetry violating superconductors. It remains open whether this type of phenomenon might become part of a building block for future dissipationless electronics.

\vspace{5mm}

\noindent
 {\Large \bf Methods}

\noindent
{\bf Fabrication of junctions:} Nb/Ru/SRO Josephson junctions were fabricated from SRO-Ru eutectic single crystal using the floating zone method~\cite{Bobowski2019}. The junctions were prepared by the following protocol. i) A 300-nm-thick SiO$_2$ layer was deposited by RF sputtering with a backing pressure of $\approx$10$^{-7}$~mbar on a polished $ab$-surface of a rectangular piece of SRO-Ru eutectic crystal with dimensions of 3$\times$3$\times$0.5 mm$^3$. ii) Using laser lithography and etching with CHF3 gas, a window was opened over a single Ru-inclusion (see Fig.~\ref{device-RT}b and Figure~S1a). During this etching process, a fluoride thin film may be generated on the surface of the sample. Therefore, before removing the photo resist, O$_2$ plasma cleaning was preformed to etch away the fluoride film. iii) A 600-nm-thick top Nb electrode was deposited using a lift-off technique. The Nb was deposited by magnetron sputtering with a base pressure of $\approx$ 10$^{-7}$ mbar.
Note that the Nb top electrode deposited by this protocol is not only in contact with Ru but also the $ab$-surface of SRO. The $ab$-surface of SRO develops a poor contact with other normal metals. It may happen because of atomic reconstruction at the surface~\cite{Veenstra2013} that avoids to develop superconducting coupling at Nb/SRO and no flow of critical current through the interface between Nb and $ab$-surface of SRO. Therefore, superconducting coupling mainly develops through Ru metal. 

\noindent
{\bf Electrical Measurements} Electronic transport properties were investigated by measuring $V$($I$) curves using a four-point current-bias technique with two contacts on the Nb electrode and two contacts on the side of the SRO crystal as shown in the schematic diagram in Fig.~\ref{device-RT}c. The current is applied using a Yokogawa current source (GS200) and the voltage is measured using a Keithley 2082. The measurements were performed with a He-3 cryostat down to 300~mK. The cryostat was magnetically shielded with high-permeability material (Hamamatsu Photonics, mu-metal). Inside the shield, we placed a superconducting magnet (with maximum field range of 1~T) to apply magnetic fields.

\vspace{5mm}

\noindent
 {\Large \bf Data availability}
 
The data that support the findings of this study are available from the corresponding authors upon reasonable request.

\vspace*{5mm}

\noindent {\Large \bf Acknowledments}

\noindent We are grateful to B. Zinkl and G. Mattoni for helpful discussion. We acknowledge the support from International Center for Materials Nanoarchitectonics (MANA) in the National Institute for Materials Science (NIMS), Japan. This work is supported by JSPS KAKENHI (Nos. JP15H05852, JP15K21717, JP22H01168, JP23K17670, JP22H04473 and JP26287078), JSPS-EPSRC Core-to-Core Programme (No. JPJSCCA20170002), and the Swiss National Science Foundation (SNSF) through Division II (No. 184739). M.S.A. and J.W.A.R. acknowledges funding from the EPSRC International Network Grant “Oxide Superspin” (No. EP/P026311/1).

\vspace*{5mm}

\noindent {\Large \bf Authors Contributions}

\noindent M.S.A. devised the experiments and performed the measurements. T.N., R.I. and M.S.A. prepared the devices. S.Y. was also involved in the measurements. M.S.A., S.A. J.W.A.R. and Y.M. analysed the data. M.S. provided theoretical input. M.S.A. wrote the paper and all the authors were involved in revision and discussion.

\vspace*{5mm}

\noindent {\Large \bf Competing interests}

\noindent The authors declare no competing interests.


\vspace{5mm}
\noindent {\Large \bf References}

\begin{thebibliography}{99}
\bibitem{Ando2020} Ando, F. et al. Observation of superconducting diode effect. {\it Nature} {\bf 584}, 373-376 (2020).

\bibitem{Narita2022} Narita, H. et al. Field-free superconducting diode effect in noncentrosymmetric superconductor/ferromagnet multilayers. {\it Nat. Nanotech.} {\bf 17}, 823-828 (2022).

\bibitem{Ye2022} Ye, C. et al. Non-reciprocal transport in a bilayer of MnBi$_2$Te$_4$ and Pt. {\it Nano Lett.} {\bf 22}, 1366-1373 (2022).

\bibitem{Wakatsuki2017} Wakatsuki, R. et al. Non-reciprocal charge transport in noncentrosymmetric superconductors. {\it Sci. Adv.} {\bf 3}, e1602390 (2017).

\bibitem{Bauriedl2022} Bauriedl, L. C. B. et al. Supercurrent diode effect and magnetochiral anisotropy in few layer NbSe${_2}$. {\it Nat. Commun.} {\bf 13},4266 (2022).

\bibitem{Hou2022} Hou, Y. et al. Ubiquitous superconducting diode effect in superconductor thin films. {\it Phys. Rev. Lett.} {\bf 131}, 027001 (2023).

\bibitem{Pal2021} Pal, B. et al. Josephson diode effect from Cooper pair momentum in a topological semimetal. {\it Nat. Phys.} {\bf 18}, 1228-1233 (2022).

\bibitem{Baumgartner2022} Baumgartner, C. et al. Supercurrent rectification and magnetochiral effects in symmetric Josephson junctions. {\it Nat. Nanotechnol.} {\bf 17}, 39-44 (2022).

\bibitem{Wu2022} Wu, H. et al. The field-free Josephson diode in a van der Waals heterostructure. {\it Nature} {\bf 604}, 653-656 (2022).

\bibitem{Jeon2022} Jeon, K-R et al., Zero-field polarity-reversible Josephson supercurrent diodes enabled by a proximity-magnetized Pt barrier. 1) {\it Nat. Mat.} {\bf 21}, 1008-1013 (2022).

\bibitem{Lin2022} Lin, J-X et al., Zero-field superconducting diode effect in small-twist-angle trilayer graphene. {\it Nat. Phys.} {\bf 18}, 1221-1227 (2022).

\bibitem{Merida2023} Díez-Mérida J. et al., Symmetry-broken Josephson junctions and superconducting diodes in magic-angle twisted bilayer graphene. {\it Nat. Commun.} {\bf 14}, 2396 (2023).

\bibitem{Rikken1997} Rikken, G. L. J. A. \& Raupach, E. Observation of magneto-chiral dichroism. {\it Nature} {\bf 390}, 493-494 (1997).

\bibitem{Rikken2001} Rikken, G. L. J. A. Folling, J. \& Wyder, P. Electrical magnetochiral anisotropy. {\it Phys. Rev. Lett.} {\bf 87}, 236602 (2001).

\bibitem{Daido2022} Daido, A., Ikeda, Y. \& Yanase, Y. Intrinsic superconducting diode effect. {\it Phys. Rev. Lett.} {\bf 128}, 037001 (2022).

\bibitem{Yuan2022} Yuan, N. F. Q. \& Fu, L. Supercurrent diode effect and finite-momentum superconductors. {\it Proc. Natl. Acad. Sci.} {\bf 119}, e2119548119 (2022).

\bibitem{Ilic2022} Ilić, S. \& Bergeret, F. S. Theory of the supercurrent diode effect in Rashba superconductors with arbitrary disorder. {\it Phys. Rev. Lett.} {\bf 128}, 177001 (2022).

\bibitem{Scammell2022} Scammell, H. D., Li. J. I. A. \& Mathias S. S. Theory of zero-field superconducting diode effect in twisted trilayer graphene. {\it 2D Mater.} {\bf 9}, 025027 (2022).

\bibitem{He2022} He, J. J., Tanaka, Y. \& Nagaosa N. A phenomenological theory of superconductor diodes. {\it New J. Phys.} {\bf 24}, 053014 (2022). 

\bibitem{Zinkl2022}  Zinkl, B., Hamamoto, K., \& Sigrist, M. Symmetry conditions for the superconducting diode effect in chiral superconductors. {\it Phys. Rev. Research} {\bf 4}, 033167 (2022).

\bibitem{Hooper2004} Hooper, J. et al. Anomalous Josephson network in the Ru-Sr$_2$RuO$_4$ eutectic system. {\it Phys. Rev. B} {\bf 70}, 014510 (2004).

\bibitem{Maeno2012}	Maeno, Y., Kittaka, S., Nomura, T., Yonezawa, S. \& Ishida, K. Evaluation of spin-triplet superconductivity in Sr$_2$RuO$_4$. {\it J. Phys. Soc. Jpn} {\bf 81}, 011009 (2012).

\bibitem{Mackenzie2017} Mackenzie, A. P., Scaffidi, T., Hicks, C. W. \& Maeno, Y. Even odder after twenty-three years: the superconducting order parameter puzzle of Sr$_2$RuO$_4$. {\it npj Quantum Materials} {\bf 2}, 40 (2017).

\bibitem{Anwar2021} Anwar, M.S. \& Robinson, R. W. A., A review of electronic transport in superconducting Sr$_2$RuO$_4$ junctions. {\it Coatings} {\bf 11}, 1110 (2021).

\bibitem{Ghosh2017} Ghosh, S. S., Xin, Y., Mao, Z. \& Manousakis, E. Interface between Sr$_2$RuO$_4$ and Ru-metal inclusion: Implications for its superconductivity. {\it Phys. Rev. B} {\bf 96}, 184506 (2017).

\bibitem{Maeno1998} Maeno, Y. et al. Enhancement of superconductivity of Sr$_2$RuO$_4$ to 3 K by embedded metallic microdomains. {\it Phys. Rev. Lett.} {\bf 81}, 3765 (1998).

\bibitem{Haka2009} Kittaka, S., Yaguchi, H., \& Maeno, Y. Large enhancement of 3-K phase superconductivity in the Sr$_2$RuO$_4$-Ru eutectic system by uniaxial pressure. {\it J. Phys. Soc. Jpn} {\bf 78}, 103705 (2009).  

\bibitem{Hicks2014} Hicks, C. W. et al. Strong increase of $T_c$ of Sr$_2$RuO$_4$ under both tensile and compressive strain. {\it Science} {\bf 344}, 283-285 (2014).

\bibitem{Taniguchi2015} Taniguchi, H., Nishimura, K., Goh, S. K., Yonezawa, S. \& Maeno, Y. Higher-$T_{\rm c}$ superconducting phase in Sr$_2$RuO$_4$ induced by in-plane uniaxial pressure. {\it J. Phys. Soc. Jpn.} {\bf 84}, 014707 (2015).

\bibitem{Hicks2017} Steppke, A. et al. Strong peak in $T_{\rm c}$ of Sr$_2$RuO$_4$ under uniaxial pressure. {\it Science} {\bf 355}, eaaf9398 (2017).

\bibitem{Mackenzie1998} Mackenzie, A. P. et al. Extremely strong dependence of superconductivity on disorder in Sr$_2$RuO$_4$. {\it Phys. Rev. Lett.} {\bf 80}, 161 (1998).

\bibitem{Luke1998} Luke, G. M. et al. Time-reversal symmetry-breaking superconductivity in Sr$_2$RuO$_4$. {\it Nature} {\bf 394}, 558-561 (1998).

\bibitem{Grinenko2021} Grinenko, V. et al. Split superconducting and time-reversal symmetry breaking transitions in Sr$_2$RuO$_4$ under stress. {\it Nat. Phys.} {\bf 17}, 748-754 (2021). 

\bibitem{Xia2006} Xia, J., Maeno, Y., Beyersdorf, P.T., Fejer, M.M., \& Kapitulnik, A. High resolution polar Kerr effect measurements of Sr$_2$RuO$_4$: Evidence for broken time-reversal symmetry in the superconducting state. {\it Phys. Rev. Lett.} {\bf 97}, 167002 (2006).

\bibitem{Kindwingira2006}	Kidwingira, F., Strand, J. D., Harlingen, D. J. V., \&  Maeno, Y. Dynamical superconducting order parameter domains in Sr$_2$RuO$_4$. {\it Science} {\bf 314}, 1267-1271 (2006).

\bibitem{Nakamura2011} Nakamura, T. et al. Topological competition of superconductivity in Pb/Ru/Sr$_2$RuO$_4$ junctions. {\it Phys. Rev. B} {\bf 84}, 060512 (2011).

\bibitem{Nakamura2012} Nakamura, T. et al. Essential configuration of Pb/Ru/Sr$_2$RuO$_4$ junctions exhibiting anomalous superconducting interference. {\it J. Phys. Soc. Jpn} {\bf 81}, 064708 (2012).

\bibitem{Anwar2013} Anwar, M. S. et al. Anomalous switching in Nb/Ru/Sr$_2$RuO$_4$ topological junctions by chiral domain wall motion. {\it Sci. Rep.} {\bf 3}, 2480 (2013).

\bibitem{Anwar2017} Anwar, M. S. et al. Multicomponent order parameter superconductivity of Sr$_2$RuO$_4$. {\it Phys. Rev. B} {\bf 95}, 224509 (2017).

\bibitem{Hassinger2017} Hassinger, E. et al. Vertical line nodes in the superconducting gap structure of Sr$_2$RuO$_4$. {\it Phys. Rev. X} {\bf 7}, 011032 (2017). 

\bibitem{Pustogow2019} Pustogow, A. et al. Constraints on the superconducting order parameter in Sr$_2$RuO$_4$ from oxygen-17 nuclear magnetic resonance. {\it Nature} {\bf 574}, 72-75 (2019). 

\bibitem{Ishida2020} Ishida, K., Manago, M., Kinjo, K., \& Maeno, Y. Reduction of the $^{17}$O Knight shift in the superconducting state and the heat-up effect by NMR pulses on Sr$_2$RuO$_4$. {\it J. Phys. Soc. Jpn} {\bf 89}, 034712 (2020).

\bibitem{Petsch2020} Petsch, A. N. et al. Reduction of the spin susceptibility in the superconducting state of Sr$_2$RuO$_4$ observed by polarized neutron scattering. {\it Phys. Rev. Lett.} {\bf 125}, 217004 (2020). 

\bibitem{Li2021} Li, Y-S. et al. High-sensitivity heat-capacity measurements on Sr$_2$RuO$_4$ under uniaxial pressure. {\it PNAS} {\bf 118}, e2020492118 (2021). 

\bibitem{Romer2019} Romer, A. T., Scherer, D. D., Eremin, I. M., Hirschfeld, P. J. \& Andersen, B. M. Knight shift and leading superconducting instability from spin fluctuations in Sr$_2$RuO$_4$. {\it Phys. Rev. Lett.} {\bf 123}, 247001 (2019). 

\bibitem{Suh2020} Suh, H. G., Menke, H., Brydon, P. M. R., Timm, C., Ramires, A. and Agterberg, D. F. Stabilizing even-parity chiral superconductivity in Sr$_2$RuO$_4$. {\it Phys. Rev. Research} {\bf 2}, 032023(R) (2020). 

\bibitem{Kivelson2020} Kivelson, S. A., C., Y. A., Ramshaw, B. J. \& Tomale, R. A proposal for reconciling diverse experiments on the superconducting state in Sr$_2$RuO$_4$. {\it npj Quantum Mat.} {\bf 5}, 43 (2020).

\bibitem{Nago2014} Nago, Y. et al., Superconducting transition of Ru in SQUIDs with Nb/Ru/Sr$_2$RuO$_4$ junctions. {\it J. Phys.: Conf. Ser.} {\bf 568}, 022031 (2014). 

\bibitem{Kaneyasu2010-1} Kaneyazu, H. \& Sigrist, M. Nucleation of vortex state in Ru-Inclusion in eutectic ruthenium oxide Sr$_2$RuO$_4$–Ru. {\it J. Phys. Soc. Jpn} {\bf 79}, 053706 (2010).

\bibitem{Kaneyasu2010-2} Kaneyasue, H., Hayashi, N., Gut, B., Makoshi, K. \& Sigrist, M. Phase transition in the 3-Kelvin phase of eutectic Sr$_2$RuO$_4$-Ru. {\it J. Phys. Soc. Jpn} {\bf 79}, 104705 (2010).

\bibitem{Etter2014} Etter, S. B., Kaneyasu, H., Ossadnik, M. \& Sigrist, M. Limiting mechanism for critical current in topologically frustrated Josephson junctions. {\it Phys. Rev. B} {\bf 90}, 024515 (2014). 

\bibitem{Matzdorf2000} Matzdorf, R. et al. Ferromagnetism stabilized by lattice distortion at the surface of the $p$-wave superconductor Sr$_2$RuO$_4$. {\it Science} {\bf 289}, 746-748 (2000). 

\bibitem{Angelo2021} Fittipaldi, R. et al. Unveiling unconventional magnetism at the surface of Sr$_2$RuO$_4$. {\it Nat. Commun.} {\bf 12}, 5792 (2021).

\bibitem{Anwar2015} Anwar, M. S. et al. Ferromagnetic SrRuO$_3$ thin-film deposition on a spin-triplet superconductor Sr$_2$RuO$_4$ with a highly conducting interface. {\it Appl. Phys. Express} {\bf 8}, 015502 (2015).  

\bibitem{Anwar2016} Anwar, M. S. et al. Direct penetration of spin-triplet superconductivity into a ferromagnet in Au/SrRuO$_3$/Sr$_2$RuO$_4$ junctions. {\it Nat. Commun.} {\bf 7}, 13220 (2016).


\bibitem{Chahid2023} Chahid, S., Teknowijoyo, S., Mowgood, I. and Gulian, A., High-frequency diode effect in superconducting Nb$_3$Sn microbridges. {\it Phys. Rev. B} {\bf 107}, 054506 (2023).

\bibitem{Gutfreund2023} Gutfreund, A. et al., Direct observation of a superconducting vortex diode. {\it Nat.  Commun.} {\bf 14}, 1630 (2023).

\bibitem{Romanenko2014} Romanenko, A., Grassellino, A., Melnychuk, O., Sergatskov, D. A. Dependence of the residual surface resistance of superconducting radio frequency cavities on the cooling dynamics around $T_{\rm c}$. {\it J. App. Phys.} {\bf 115}, 184903 (2014).

\bibitem{Kubo2016} Kubo, T., Flux trapping in superconducting accelerating cavities during cooling down with a spatial temperature gradient {\it Prog. Theor. Exp. Phys.} 053G01 (2016).

\bibitem{Bobowski2019} Bobowski, J. S. et al. Improved single-crystal growth of Sr$_2$RuO$_4$. {\it Condens. Matter} {\bf 4}, 6 (2019).

\bibitem{Veenstra2013} Veenstra, C. N. et al. Determining the surface-to-bulk progression in the normal-state electronic structure of Sr$_2$RuO$_4$ by angle-resolved photoemission and density functional theory. {\it Phys. Rev. Lett.} {\bf 110}, 0977004 (2013).


\end{thebibliography}
\end{document}